ICTM 5/2017

Jolanta Kowal (Ed.)

Anna Kuzio (Ed.)

Juho Mäkiö (Ed.)

Grażyna Paliwoda-Pękosz (Ed.)

Piotr Soja (Ed.)

Ralph Sonntag (Ed.)

# ICT Management for Global Competitiveness and Economic Growth in Emerging Economies (ICTM)

**International Conference on ICT Management for Global Competitiveness and Economic Growth in Emerging Economies**

**Wrocław, Poland, October 23-24, 2017**

**Proceedings**

# Improving the quality of healthcare through Internet of Things

by

**Cornel Turcu**, Stefan cel Mare University of Suceava, Romania, cturcu@eed.usv.ro
**Cristina Turcu**, Stefan cel Mare University of Suceava, Romania, cristina@eed.usv.ro

## ABSTRACT

This paper attempts to outline how the adoption of Internet of Things (IoT) in healthcare can create real economic value and improve patient experience. Thus, getting the maximum benefits requires understanding both the IoT paradigm and the enabling technologies, and how IoT can be applied in the field of healthcare. We will mention some open challenging issues to be addressed by the research community, and not only. Besides the real barriers in adopting the Internet of Things, there are some advantages regard collecting and processing patient data, and monitoring the daily health states of individuals, just to name a few. These aspects could revolutionize the healthcare industry.

**Keywords**: Internet of Things, IoT platforms, healthcare, big data, cloud computing.

## INTRODUCTION

In recent years, a transformative paradigm known as Internet of Things (IoT) is coming in our lives. According to (Lucero & others, 2016) the IoT "is a technology concept that is currently transforming and redefining virtually all markets and industries in fundamental ways". Among the types of services that stand to benefit from Internet of Things technology, healthcare is one of the most promising. In order to benefit from the full potential of Internet of Things, we need to understand where real value can be created; we also need to successfully tackle challenges.

Over time, several definitions for IoT have been provided worldwide, but currently, none of them have been universally accepted. Nevertheless, these different definitions reflect various perspectives and support different business interests and their analysis can help us understand the Internet of Things phenomenon.

Despite the fact that the list is not complete and will certainly be extended in the future, we can mention some definitions considered relevant by scientific literature.

**Table 1.** Definitions of the Internet of Things

| IoT Definition | By |
|---|---|
| "The term 'Internet of Things' (IoT) denotes a trend where a large number of embedded devices employ communication services offered by the Internet protocols. Many of these devices, often called 'smart objects,' are not directly operated by humans, but exist as components in buildings or vehicles, or are spread out in the environment." | (Tschofenig, Arkko, Thaler, & McPherson, 2015) |
| "A network of items — each embedded with sensors — which are connected to the Internet." | (IEEE Institute, 2014) |
| "The Internet of Things (IoT) is the network of physical objects that contain embedded technology to communicate and sense or interact with their internal states or the external environment." | ("Internet of Things Defined - Tech Definitions by Gartner," n.d.) |
| "The IoT creates an intelligent, invisible network fabric that can be sensed, controlled and programmed. IoT-enabled products employ embedded technology that allows them to communicate, directly or indirectly, with each other or the Internet." | (Chase, 2013) |
| Interconnection of sensing and actuating devices providing the ability to share information across platforms through a unified framework, developing a common operating picture for enabling innovative applications. This is achieved by seamless large scale sensing, data analytics and information representation using cutting edge ubiquitous sensing and cloud computing. | (Gubbi, Buyya, Marusic, & Palaniswami, 2013) |

Most Internet of Things definitions focus on common aspects, such as the interconnection of uniquely identifiable things and the connection of things to the Internet, big data, intelligence, etc. There is also a general agreement on the fact that IoT changes the way things are perceived and used in the smartest way possible, in order to meet the individial's needs to the best possible extent. For example, patients with heart or respiratory diseases or diabetes could be monitored through ingestible or attached sensors. These devices can transmit readings and alert the patient, nurses and doctors when vital signs indicate an imminent problem; the purpose is to take action in order to avoid crises, and also unnecessary hospitalization.

According to a survey conducted by Forrester Consulting on behalf of Zebra Technologies ("IoT to Revolutionize Healthcare Industry: Survey," 2015), 97% of the surveyed healthcare industry professionals agree that "the IoT is the most strategic solution their organization will undertake this decade". Nine of ten healthcare IT departments are prepared to make the necessary changes in order to implement IoT solutions. Moreover, the survey reveals that over half of the healthcare respondents have already begun implementing IoT solutions in their practice.

By analyzing various worldwide studies and surveys, we can ascertain that over the last years, the interest and intention to invest in an IoT-based solution has steadily increased. Just a few years ago, some companies didn't even know what IoT was. And now there is a belief that for any new societal challenge, there is always an IoT-based solution that successfully addresses it (Atzori, Iera, & Morabito, 2017). Nevertheless, IoT is proposed almost anytime and anywhere, as a panacea of the Information Communication Technology (ICT) world and we cannot but ask ourselves if this is hype or reality. But only the future will tell. Meanwhile, industrial players are taking advantage of this impressive increase in the market momentum and use the popularity of IoT as a strong brand for consumer-oriented technology solutions (Atzori, Iera, & Morabito, 2017).

It cannot be denied that IoT could provide many benefits in different fields, including healthcare. And this is confirmed by the numerous results obtained in research activities. Thus, over the past years, the number of publications related to IoT in healthcare grew noticeably. Nevertheless, it should be noted that scientific literature does not always contribute to clarifying problems, as is the case with IoT definitions, which sometimes are inconsistent with each other. In order to explain what IoT refers to, several surveys have been published worldwide, each of them focusing on certain aspects: things (Atzori, Iera, & Morabito, 2010), architecture (Gubbi, Buyya, Marusic, & Palaniswami, 2013) (Weyrich & Ebert, 2016), standards (Stackowiak, Licht, Mantha, & Nagode, 2015), enabling technologies (Al-Fuqaha, Guizani, Mohammadi, Aledhari, & Ayyash, 2015), platforms, challenges (Miorandi, Sicari, De Pellegrini, & Chlamtac, 2012), applications (Man, Na, & Kit, 2015), etc.

This paper attempts to outline how the adoption of IoT in healthcare can create real economic value and improve patient experience (PX). Thus, getting the maximum benefits requires an understanding of the IoT paradigm and the enabling technologies, and how they can be applied in the healthcare domain. Also, the paper presents the key issues that remain to be tackled.

The rest of the paper is organized as follows. Section II overviews both the positive and negative aspects IoT has ever the field of healthcare. Some of the main enabling technologies, such as big data, cloud computing, etc. that allow the expansion of power and reach of information are presented in Section III. Some of the various barriers that hamper the wider uptake of IoT in healthcare are presented in IV. Section V and VI introduce some existing IoT platforms and IoT-based applications. Section VII briefs on a few future research directions. The last section concludes the paper.

Currently, it is difficult to estimate how the healthcare field will benefit from the adoption of IoT and which will be the real financial benefits resulting from these new insights. In the next subsection, we present some of the worldwide studies and surveys that estimate the economic impact of adopting IoT in healthcare.

## IMPACT

The goals of applying IoT in healthcare support the key objectives of the Digital Agenda for Europe: "improving the quality of healthcare, reducing medical costs and fostering independent living for those needing care" (eHealth Action Plan 2012-2020-Innovative healthcare for the 21st century, 2012).

(Manyika et al., 2015) estimates a higher potential value for IoT in advanced economies over the next ten years, due to higher value per use. Moreover, there is a high potential for Internet of Things in developing economies. Thus, according to the same report, nearly 40 percent of the value could be generated in the developing economies.

According to a report from MarketResearch.com, the healthcare Internet of Things market segment is poised to hit $117 billion by 2020 ("Big Data in Internet of Things (IoT): Key Trends, Opportunities and Market Forecasts 2015 – 2020," n.d.). A McKinsey report estimates that by 2025, the economic impact of the Internet of Things (IoT) will be between $3.9 and $11 trillion dollars a year, equivalent to about 11 percent of the world economy (exhibit) (Manyika et al., n.d.). In (Manyika et al., 2013), cloud computing (and related trends such as big data, and the Internet of Things) is projected to have a "collective economic impact" of between $10–20 trillion annually in 2025. By 2025, healthcare applications and related IoT-based services, such as mobile health (m-Health) and telecare (that enable medical wellness, prevention, diagnosis, treatment and monitoring services to be delivered efficiently through electronic media) are expected to create about $1.1–$2.5 trillion in growth, on an annual basis. It is the biggest economic impact on a global scale, taking into account that the whole annual economic impact caused by IoT is estimated to range from $2.7 trillion to $6.2 trillion by 2025 (Manyika et al., 2013).

It should be noted that, in addition and in relation to the direct effects of adopting IoT technologies in various areas, an entire dynamic industry is evolving. IoT creates new opportunities for both incumbents and new players. For instance, producers of medical devices are creating new business models by using IoT links and data, in order to offer their products as a service.

In "The Patient Will See You Now" (Topol, 2015), Eric Topol, one of the nation's top physicians, quoting an article in MIT Technology Review (Cutler, 2013), argues that the patient is the "single most unused person in health care". The author highlights how the delivery of healthcare and related services from health sciences (pharma and devices) is influenced, among other things, by the developments of information technology.

Also, according to (Cutler, 2013), the role of the patient should be re-imagined, so that the patient becomes a participant and a contributor to healthcare.

Currently, there are numerous mobile health devices and software applications for mobile devices that offer information related to the owner's health status.

Giving patients the chance to monitor their own health could change the way people perceive themselves, their illness, and the people who care for them (Cutler, 2013). Patients play an active role in their own care and this can be accomplished by granting them access to their personal health records and by empowering them to own their personal medical data. Therefore, integrating user-generated data with official medical data enables integrated and personalized healthcare. Despite the fact that this approach proves useful to patients, some doctors are hesitant to apply it in reality.

According to various research teams (McKinsey, 2011), healthcare providers, for instance, discard 90 percent of the data they generate. Instead, they should convert the terabytes and zettabytes of big data that is not currently used (classified as dark data) into useful data in order to improve patient experience. Moreover, both patients and physicians must be willing and able to use insights from this data. But this is possible only with adequate technologies (Manyika et al., 2015).

The following significant changes in technology have come together to enable the rise of IoT.

**ENABLING TECHNOLOGIES**

Over the past several years, the development of various technologies has enabled the evolution of the Internet of Things. Thus, the progress of the IoT concept involves bringing together RFID, connectivity, cloud (Abawajy & Hassan, 2017) (Atlam, Alenezi, Alharthi, Walters, & Wills, 2017), big data analytics (Papadokostaki et al., 2017) (Bhatt, Dey, & Ashour, 2017), and application development capabilities, just to name a few. Related to healthcare, we could mention the development of the IoT infrastructure in this field, which includes home monitoring and remote care, such as telemedicine, smart wearables, etc. (figure 1).

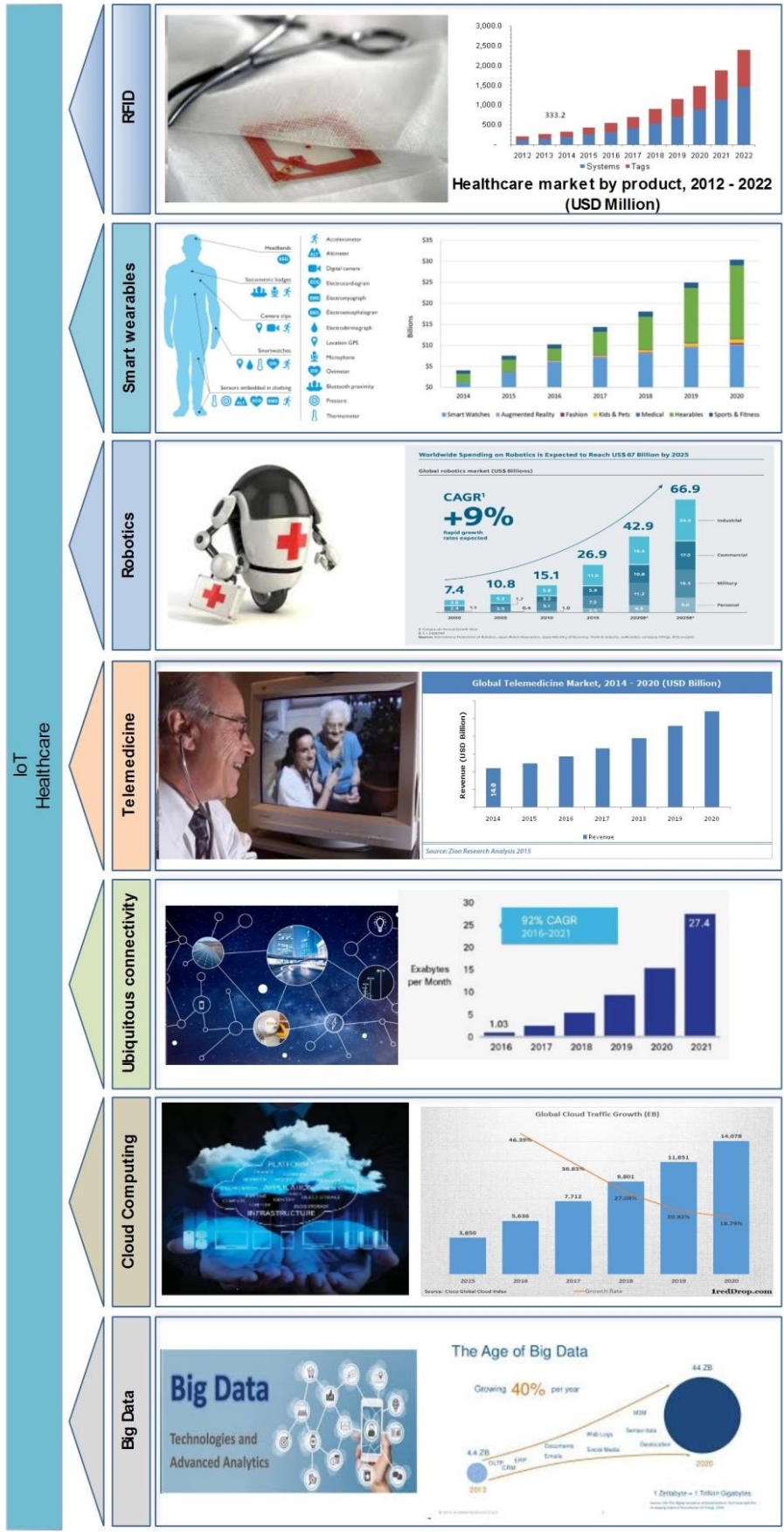

Figure 1. Enabling technologies for IoT Healthcare.

The millions and millions of things connected to IoT produce huge volumes of data. In these circumstances, an efficient, scalable and accessible way is needed in order to handle securely all this information and to produce value. Cloud technologies enable a solution for securely storing, processing and analyzing data, especially large data.

In order to monetize data and extract meaning, it does not suffice to collect or access larger sets of data. There is also a real need for analyzing and mining large amounts of data for the benefit of citizens, researchers, practitioners, businesses and decision makers (*eHealth Action Plan 2012-2020-Innovative healthcare for the 21st century*, 2012). Hence, the wealth and complexity of data and information collected from an increased number of different devices, systems and sensors, makes it essential to adopt big data technology in order to store, analyze, search, share, view query and update it. To allow faster and smarter decisions-making required for high performance and competitive advantage, machine learning, a branch of artificial intelligence, is used to add an intelligence layer to big data. Thus, in order to harness true benefits, big data is turned into smart data. But some healthcare scenarios need real-time data in order to enable doctors to give recommendations to the patients who need them. So, fast data is needed to provide instant results and responses.

For example, in the case of patients with heart diseases, doctors can develop predictive scenarios based on the patient's various biological parameters, and they can figure out how to intervene before a heart attack occurs. Moreover, chronic disease management can also benefit from the implementation of IoT-based solutions.

The application of IoT in healthcare will be accelerated by the convergence of the above mentioned technologies and not only.

Nevertheless, the implementation of the Internet of Things in various fields, such as healthcare, is raising concerns about data privacy and security. Worldwide, numerous studies reveal the need for greater focus on cybersecurity in order to protect patient data. Several scientific papers and reports that diagnose cyber threats and identify mitigation techniques and good practices for healthcare industry, e.g. ("Cyber security and resilience for Smart Hospitals — ENISA," 2016) have been published.

The U.S. Food and Drug Administration offers guidelines for medical devices (Health, n.d.) ("Postmarket Management of Cybersecurity in Medical Devices. Guidance for Industry and Food and Drug Administration Staff," 2016) and regulators regulate connected devices used by patients.

By adopting new specific legislation, governments are making a major step forward in the widespread adoption of the Internet of Things, but they should also remain very cautious and rigorous regarding the potential misuses of IoT technologies.

Internet of Things provides new and great opportunities, but also new and great challenges, some of them being illustrated in the next subsection.

## CHALLENGES

The adoption of IoT in healthcare faces several challenges. This shift from traditional healthcare to using new technology and engineering innovations requires healthcare reforms & mandates. The lack of IoT skills and knowledge among employees and management is an important challenge that healthcare providers need to address. Moreover, there is a general lack of confidence in IoT solutions among patients, citizens and healthcare professionals, that might be generated by certain lack of transparency regarding the use of data collected from various devices, such as mobile devices, sensors, etc. (*eHealth Action Plan 2012-2020-Innovative healthcare for the 21st century*, 2012).

The Internet of Things will definitely change the way both doctors and patients perceive healthcare. Thus, as we mentioned before, the role of the patient needs to be re-imagined.

Currently, software applications, including those for mobile devices, could provide information, various possibilities of 'self-quantification' and could even prove to be efficient diagnostic tools. In fact, they represent new ways of care, which reduces the distinction between the traditional provision of clinical care by doctors, and the self-administration of care by the patient. Therefore, the issue revolves around defining the roles that healthcare professionals, but also software and hardware developers, equipment suppliers, and so on, could play in the value chain of IoT healthcare.

Besides, there are also technological challenges. One of them, raising concerns, is the lack of widely accepted standards (Jabeur & Haddad, 2015). There are serious interoperability problems due to the heterogeneity of hardware/ software specifications and capabilities, the inconsistency of data representations, and their random mobility, the use of various medical devices, sensors, and patient own-devices.

In order to meet the specific requirements, it is necessary to extend and/or revise the existing standards for storing, viewing and sharing data.

Also, additional challenges concerning trust, privacy, and security need to be addressed in respect to making the adoption of IoT a reality in healthcare. Effective data protection is vital for building trust in IoT. According to (Contu, Middleton, Perkins, & Akshay, 2016) "the effort of securing IoT is expected to focus more and more on the management, analytics and provisioning of devices and their data". The importance given to this subject is highlighted, among other things, by the estimated spending. Thus, the worldwide spending on the Internet of Things security will reach $348 million in 2016, a 23.7 percent increase from 2015, when the numbers mounted up to $281.5 million, according to (Contu, Middleton, Perkins, & Akshay, 2016). The same report estimates that in 2018, the costs will reach $547 million (see Table 2).

**Table 2.** Worldwide IoT Security Spending Forecast (Millions of Dollars) (Contu, Middleton, Perkins, & Akshay, 2016)

| 2014 | 2015 | 2016 | 2017 | 2018 |
|------|------|------|------|------|
| 231.86 | 281.54 | 348.32 | 433.95 | 547.2 |

Artificial intelligence techniques, which have proven their performance in addressing the issues of extremely dynamic, uncertain and heterogeneous environments, could bring solutions to some issues in the field of IoT, such as data processing and even security problems. At present, there are already some papers published addressing the integration of such techniques in the context of IoT, with applicability in various domains, e.g. (Rapti, Karageorgos, Houstis, & Houstis, 2017) (do Nascimento & de Lucena, 2017) (Singh & Chopra, 2017) (Mzahm, Ahmad, & Tang, 2013) (Mzahm, Ahmad, Tang, & Ahmad, 2016) (Kortuem, Kawsar, Sundramoorthy, & Fitton, 2010). Also, we could mention some publications analyzing various solutions provided by artificial intelligence to issues related to IoT, specific to the field of healthcare (Qi et al., 2017) (Korzun, 2017) (Vargiu & Zambonelli, 2017).

**INTERNET OF THINGS PLATFORMS**

Internet of Things platforms are a key tools in addressing and redressing the aforementioned problems.

One of the main purposes of IoT platforms is to provide a solution to the increasing demand of IoT applications in various domains. Thus, these platforms enable the IoT developers and implementers to focus on the specific, differentiated and unique value the application provides and "outsource common, industry-wide features and functionality" (Lucero & others, 2016). In fact, these IoT platforms are dedicated software suites that offer a full spectrum of functional capabilities (Natis, Lheureux, Thomas, Pezzini, & Velosa, 2015).

The many versions of the IoT platforms vary widely in functionality. At present, there are numerous IoT platforms; some of these platforms that have been applied in healthcare are mentioned in Table 3; nevertheless, neither one of them is adequate to support the end-to-end workflow of an advanced IoT solution.

**Table 3.** IoT Platforms

| Platform | Producer | Ref |
|----------|----------|-----|
| Aeris | Aeris | ("IoT Services and Healthcare Focus on the Patient \| Aeris," n.d.) |
| AllJoyn | AllSeen Alliance | ("AllSeen Alliance," n.d.) |
| General Electric's Predix | General Electric (GE) | ("Cloud-based Platform-as-a-Service (PaaS) \| Predix.io," n.d.) |
| HealthSaaS | HealthSaaS, Inc. | ("HealthSaaS \| The Internet of Things (IoT) Platform for Healthcare," n.d.) |
| IoTivity | Open Connectivity Foundation, Inc | ("OCF - Healthcare," n.d.) (Lee, Jeon, & Kim, 2016) |
| Kaa | KAAIOT | ("IoT Healthcare Solutions - Medical Internet of Things for Healthcare Devices and Hospitals," n.d.) |
| Kore | Kore Inc. | ("M2M Healthcare Applications," n.d.) |
| Telit | Telit | ("Healthcare – Telit," n.d.) |
| ThingWorx | PTC | ("Enterprise IoT Platform," n.d.) |
| Xively | LogMeIn | ("IoT Platform for Medical Devices \| Xively by LogMeIn," n.d.) |

The list is open and at any moment, new producers or early market leaders could provide new alternatives.

The potential of IoT platforms enable the development of a large number of applications. In the following subsection, we present a range of applications for the field of healthcare and also some futuristic applications.

## APPLICATIONS

At present, the Internet of Things has moved beyond concepts and trials and a range of industries begun to benefit from it (Geschickter, 2016).

Applications resulting from the adoption of the Internet of Things in healthcare are numerous and diverse. They can be grouped into various categories, such as:

- tracking of objects and people (staff and patients); identification and authentication of people; automatic data collection and sensing (Atzori, Iera, & Morabito, 2010) (Vilamovska et al., 2009); disease diagnosis applications;

- single-condition applications (refer to a specific disease or infirmity) and clustered-condition applications (deal with a number of diseases or conditions together as a whole) (Islam, Kwak, Kabir, Hossain, & Kwak, 2015).

The IoT-based applications lead to what is called the "quantified self" movement, allowing people to get highly involved in healthcare by using different devices and sensors, mobile phones, etc.; for example, there are blood pressure or ECG monitors or devices that monitor sleep activity, even ingestible sensors which can transmit information to doctors.

Thereby, new opportunities occur in managing health and disease aspects.

## FUTURE RESEARCH DIRECTIONS

In order to recast and revolutionize patient healthcare on a worldwide level, other trends take into account Web of Things (WoT) and the shift of the WoT towards a Social Web of Things (SWoT) (Zeng, Guo, & Cheng, 2011) (Chung et al., 2013) (Jabeur & Haddad, 2015) or even Social Web of Intelligent Things (Console, Lombardi, Picardi, & Simeoni, 2011). The WoT paradigm implies the use of web protocols and technologies in order to integrate and connect things (various real-world living or non-living entities) that become part of the World Wide Web. The Social Web of Things concept involves the use of social networks. One research trend treats Social Web of Intelligent Things (SWIT), as "an evolution of both the 'Web of Things' and 'Smart Objects' paradigms. In a SWIT, things become entities capable of an intelligent and social behavior" (Console, Lombardi, Picardi, & Simeoni, 2011).

But so far, in the current literature, there aren't many publications dealing with these approaches, especially regarding the field of healthcare. This can be explained by the many challenges as well as the lack of maturity of some of the technologies involved.

**CONCLUSION**

The emerging development of Internet of Things is expected to provide solutions across a wide range of areas, including healthcare. Currently, more and more companies in this field recognize the transformational role of IoT solutions. In recent years, the momentum for IoT solutions in healthcare, and also in other domains, has increased and it is expected to be further facilitated by the development of enabling technologies, some of which presented in this paper. Thus, in order to determine better outcomes, enable faster decisions, and increase autonomous decision making various technologies need to be used, such as big, smart, and fast data, cloud computing, etc. The way these technologies and the emergence of new ones will impact the healthcare business strategy should be carefully considered. In this paper, we highlighted some of the challenges of adopting IoT in healthcare. Thus, for instance, the IoT development is plagued by cybersecurity and data privacy issues that will present major challenges for the widespread adoption of IoT. In time, these barriers will be overcome through rapid innovations brought by technological developments.

IoT platforms, of which the most popular have been presented in this paper, play an important role in enabling the development and implementation of IoT applications. IoT-based applications could improve the delivery of healthcare services in a time-saving and low-cost manner, which will also be reflected in the potential economic impact of IoT technology in healthcare.

**REFERENCES**

Abawajy, J. H., & Hassan, M. M. (2017). Federated Internet of Things and Cloud Computing Pervasive Patient Health Monitoring System. *IEEE Communications Magazine*, *55*(1), 48–53. https://doi.org/10.1109/MCOM.2017.1600374CM

Al-Fuqaha, A., Guizani, M., Mohammadi, M., Aledhari, M., & Ayyash, M. (2015). Internet of Things: A Survey on Enabling Technologies, Protocols, and Applications. *IEEE Communications Surveys Tutorials*, *17*(4), 2347–2376. https://doi.org/10.1109/COMST.2015.2444095

AllSeen Alliance. (n.d.). Retrieved September 12, 2017, from https://allseenalliance.org/

Atlam, H., Alenezi, A., Alharthi, A., Walters, R., & Wills, G. (2017). Integration of cloud computing with internet of things: challenges and open issues (pp. 1–6). Retrieved from https://eprints.soton.ac.uk/410989/

Atzori, L., Iera, A., & Morabito, G. (2010). The Internet of Things: A survey. *Computer Networks*, *54*(15), 2787–2805. https://doi.org/10.1016/j.comnet.2010.05.010

Atzori, L., Iera, A., & Morabito, G. (2017). Understanding the Internet of Things: definition, potentials, and societal role of a fast evolving paradigm. *Ad Hoc Networks*, *56*, 122–140.

Bhatt, C. M., Dey, N., & Ashour, A. (2017). *Internet of Things and Big Data Technologies for Next Generation Healthcare* (Vol. 23). Springer.

Big Data in Internet of Things (IoT): Key Trends, Opportunities and Market Forecasts 2015 – 2020. (n.d.). Retrieved September 6, 2017, from https://www.marketresearch.com/Mind-Commerce-Publishing-v3122/Big-Data-Internet-Things-IoT-8926222/

Chase, J. (2013). The evolution of the Internet of Things. *Texas Instruments*. Retrieved from http://www.ti.com/ww/en/connect_more/pdf/SWB001.pdf

Chung, T. Y., Mashal, I., Alsaryrah, O., Huy, V., Kuo, W. H., & Agrawal, D. P. (2013). Social Web of Things: A Survey. In *2013 International Conference on Parallel and Distributed Systems* (pp. 570–575). https://doi.org/10.1109/ICPADS.2013.102

Cloud-based Platform-as-a-Service (PaaS) | Predix.io. (n.d.). Retrieved September 15, 2017, from https://www.predix.io/

Console, L., Lombardi, I., Picardi, C., & Simeoni, R. (2011). Toward a Social Web of Intelligent Things. *AI Commun.*, *24*(3), 265–279.

Contu, R., Middleton, P., Perkins, E., & Akshay, L. (2016, April 7). Forecast: IoT Security, Worldwide, 2016. Retrieved September 12, 2017, from https://www.gartner.com/doc/3277832/forecast-iot-security-worldwide-

Cutler, D. M. (2013). Why Health Care Will Become More Like Online Retail. Retrieved September 1, 2017, from https://www.technologyreview.com/s/518906/why-medicine-will-be-more-like-walmart/

Cyber security and resilience for Smart Hospitals — ENISA. (2016, November).
[Report/Study]. Retrieved September 13, 2017, from
https://www.enisa.europa.eu/publications/cyber-security-and-resilience-for-smart-
hospitals

do Nascimento, N. M., & de Lucena, C. J. P. (2017). FIoT: An agent-based framework for
self-adaptive and self-organizing applications based on the Internet of Things.
*Information Sciences*, *378*, 161–176. https://doi.org/10.1016/j.ins.2016.10.031

*eHealth Action Plan 2012-2020-Innovative healthcare for the 21st century*. (2012).
EUROPEAN COMMISSION. Retrieved from
http://ec.europa.eu/newsroom/dae/document.cfm?doc_id=4188

Enterprise IoT Platform. (n.d.). Retrieved September 15, 2017, from
https://www.thingworx.com/platforms/

Geschickter, C. (2016). Harness IoT Innovation to Generate Business Value. Retrieved
August 16, 2017, from https://www.gartner.com/doc/3479741/harness-iot-innovation-
generate-business

Gubbi, J., Buyya, R., Marusic, S., & Palaniswami, M. (2013). Internet of Things (IoT): A
vision, architectural elements, and future directions. *Future Generation Computer
Systems*, *29*(7), 1645–1660.

Health, C. for D. and R. (n.d.). Guidance Documents (Medical Devices and Radiation-
Emitting Products) - Guidance for Industry - Cybersecurity for Networked Medical
Devices Containing Off-the-Shelf (OTS) Software [WebContent]. Retrieved September
6, 2017, from
https://www.fda.gov/MedicalDevices/DeviceRegulationandGuidance/GuidanceDocume
nts/ucm077812.htm

Health, C. for D. and R. (n.d.). Guidance Documents (Medical Devices and Radiation-
Emitting Products) - Information for Healthcare Organizations about FDA's
[WebContent]. Retrieved September 6, 2017, from
https://www.fda.gov/MedicalDevices/DeviceRegulationandGuidance/GuidanceDocume
nts/ucm070634.htm

Healthcare – Telit. (n.d.). Retrieved August 22, 2017, from https://www.telit.com/industries-
solutions/healthcare/

HealthSaaS | The Internet of Things (IoT) Platform for Healthcare. (n.d.). Retrieved
    September 1, 2017, from https://www.healthsaas.net/

IEEE Institute. (2014). *Special Report: The Internet of Things*. Retrieved from
    http://theinstitute.ieee.org/static/special-report-the-internet-of-things

Internet of Things Defined - Tech Definitions by Gartner. (n.d.). Retrieved July 18, 2017,
    from https://www.gartner.com/it-glossary/internet-of-things/

IoT Healthcare Solutions - Medical Internet of Things for Healthcare Devices and Hospitals.
    (n.d.). Retrieved August 18, 2017, from https://www.kaaproject.org/healthcare/

IoT to Revolutionize Healthcare Industry: Survey. (2015), p.
    http://www.machinetomachinemagazine.com/2015/04/14/iot-to-revolutionize-
    healthcare-industry-zebra-survey/.

IoT Platform for Medical Devices| Xively by LogMeIn. (n.d.). Retrieved May 3, 2017, from
    https://www.xively.com/iot-platform-for-medical-devices

IoT Services and Healthcare Focus on the Patient | Aeris. (n.d.). Retrieved September 12,
    2017, from http://www.aeris.com/solutions/healthcare

Islam, S. M. R., Kwak, D., Kabir, M. H., Hossain, M., & Kwak, K. S. (2015). The Internet of
    Things for Health Care: A Comprehensive Survey. *IEEE Access*, *3*, 678–708.
    https://doi.org/10.1109/ACCESS.2015.2437951

Jabeur, N., & Haddad, H. (2015). From Intelligent Web of Things to Social Web of Things.
    *Facta Universitatis, Series: Electronics and Energetics*, *29*(3), 367–381.

Kortuem, G., Kawsar, F., Sundramoorthy, V., & Fitton, D. (2010). Smart objects as building
    blocks for the Internet of things. *IEEE Internet Computing*, *14*(1), 44–51.
    https://doi.org/10.1109/MIC.2009.143

Korzun, D. G. (2017). Internet of Things Meets Mobile Health Systems in Smart Spaces: An
    Overview. In *Internet of Things and Big Data Technologies for Next Generation
    Healthcare* (pp. 111–129). Springer, Cham. https://doi.org/10.1007/978-3-319-49736-
    5_6

Lee, J. C., Jeon, J. H., & Kim, S. H. (2016). Design and implementation of healthcare
    resource model on IoTivity platform. In *2016 International Conference on Information*

and Communication Technology Convergence (ICTC)* (pp. 887–891). https://doi.org/10.1109/ICTC.2016.7763322

Lucero, S., & others. (2016). IoT platforms: enabling the Internet of Things. *HIS Technology. Retrieved from https://cdn.ihs.com/www/pdf/enabling-iot.pdf*.

M2M Healthcare Applications: M2M Remote Patient Monitoring Solutions & More – KORE Telematics. (n.d.). Retrieved August 12, 2017, from http://www.korewireless.com/industry-solutions/healthcare

Man, L. C. K., Na, C. M., & Kit, N. C. (2015). IoT-based asset management system for healthcare-related industries. *International Journal of Engineering Business Management*, *7*, 19.

Manyika, J., Chui, M., Bisson, P., Woetzel, J., Dobbs, R., Bughin, J., & Aharon, D. (2015). *The Internet of Things: Mapping the value beyond the hype*. McKinsey Global Institute.

Manyika, J., Chui, M., Bisson, P., Woetzel, J., Dobbs, R., Bughin, J., & Aharon, D. (n.d.). Unlocking the potential of the Internet of Things | McKinsey & Company. Retrieved September 5, 2017, from http://www.mckinsey.com/business-functions/digital-mckinsey/our-insights/the-internet-of-things-the-value-of-digitizing-the-physical-world

Manyika, J., Chui, M., Bughin, J., Dobbs, R., Bisson, P., & Marrs, A. (2013). Disruptive technologies: Advances that will transform life, business, and the global economy | McKinsey & Company. Retrieved July 15, 2017, from http://www.mckinsey.com/business-functions/digital-mckinsey/our-insights/disruptive-technologies

McKinsey, B. D. (2011). *Big data: The next frontier for innovation, competition, and productivity*. Retrieved from McKinsey Global Institute Report website: http://www.mckinsey.com/business-functions/digital-mckinsey/our-insights/big-data-the-next-frontier-for-innovation

Miorandi, D., Sicari, S., De Pellegrini, F., & Chlamtac, I. (2012). Internet of things: Vision, applications and research challenges. *Ad Hoc Networks*, *10*(7), 1497–1516.

Mzahm, A. M., Ahmad, M. S., & Tang, A. Y. C. (2013). Agents of Things (AoT): An intelligent operational concept of the Internet of Things (IoT). In *2013 13th International Conference on Intellient Systems Design and Applications* (pp. 159–164). https://doi.org/10.1109/ISDA.2013.6920728

Mzahm, A. M., Ahmad, M. S., Tang, A. Y., & Ahmad, A. (2016). Towards a Design Model
    for Things in Agents of Things. In *Proceedings of the International Conference on
    Internet of things and Cloud Computing* (p. 41). ACM.

Natis, Y., Lheureux, B., Thomas, A., Pezzini, M., & Velosa, A. (2015). The Platform
    Architect's Guide to Designing IoT Solutions. Retrieved September 2, 2017, from
    https://www.gartner.com/doc/3153929/platform-architects-guide-designing-iot

OCF - Healthcare. (n.d.). Retrieved September 22, 2017, from
    https://openconnectivity.org/business/markets/healthcare

Papadokostaki, K., Mastorakis, G., Panagiotakis, S., Mavromoustakis, C. X., Dobre, C., &
    Batalla, J. M. (2017). Handling Big Data in the Era of Internet of Things (IoT). In
    *Advances in Mobile Cloud Computing and Big Data in the 5G Era* (pp. 3–22). Springer.

Postmarket Management of Cybersecurity in Medical Devices. Guidance for Industry and
    Food and Drug Administration Staff. (2016, December). Retrieved from
    https://www.fda.gov/downloads/MedicalDevices/DeviceRegulationandGuidance/Guida
    nceDocuments/UCM482022.pdf

Qi, J., Yang, P., Min, G., Amft, O., Dong, F., & Xu, L. (2017). Advanced internet of things
    for personalised healthcare systems: A survey. *Pervasive and Mobile Computing*, *41*,
    132–149. https://doi.org/10.1016/j.pmcj.2017.06.018

Rapti, E., Karageorgos, A., Houstis, C., & Houstis, E. (2017). Decentralized service discovery
    and selection in Internet of Things applications based on artificial potential fields.
    *Service Oriented Computing and Applications*, *11*(1), 75–86.
    https://doi.org/10.1007/s11761-016-0198-1

Singh, M. P., & Chopra, A. K. (2017). The Internet of Things and Multiagent Systems:
    Decentralized Intelligence in Distributed Computing. In *2017 IEEE 37th International
    Conference on Distributed Computing Systems (ICDCS)* (pp. 1738–1747).
    https://doi.org/10.1109/ICDCS.2017.304

Stackowiak, R., Licht, A., Mantha, V., & Nagode, L. (2015). Internet of things standards. In
    *Big Data and the Internet of Things* (pp. 185–190). Springer.

Topol, E. J. (2015). *The patient will see you now: the future of medicine is in your hands*.
    Tantor Media.

Tschofenig, H., Arkko, J., Thaler, D., & McPherson, D. (2015). *Architectural considerations in smart object networking*. Retrieved from https://tools.ietf.org/pdf/rfc7452.pdf

Vargiu, E., & Zambonelli, F. (2017). Agent abstractions for engineering IoT systems: A case study in smart healthcare. In *2017 IEEE 14th International Conference on Networking, Sensing and Control (ICNSC)* (pp. 667–672). https://doi.org/10.1109/ICNSC.2017.8000170

Vilamovska, A., Hattziandreu, E., Schindler, R., Van Oranje, C., De Vries, H., & Krapelse, J. (2009). Rfid application in healthcare–scoping and identifying areas for rfid deployment in healthcare delivery. *RAND Europe, February*.

Weyrich, M., & Ebert, C. (2016). Reference Architectures for the Internet of Things. *IEEE Software*, *33*(1), 112–116. https://doi.org/10.1109/MS.2016.20

Zeng, D., Guo, S., & Cheng, Z. (2011). The web of things: A survey. *Journal of Communications*, *6*(6), 424–438.